\documentclass[12pt]{article}

\def\Journal#1#2#3#4{{#1} {\bf #2} (#4) #3 }

\begin{document}
\title{Approximation properties of basis functions
in variational tree body problem\footnote{Talk given at the International
Conference "Quantization, Gauge Theory, and Strings" dedicated to the
memory of Professor Efim Fradkin. Moscow. 5 -- 10 May 2000. To be published
in the Conference Proceedings.}}
\author{Vladimir S. Vanyashin\footnote{e-mail: vvanyash@ff.dsu.dp.ua
~~vanyashv@heron.itep.ru}\\
{\it Dnepropetrovsk State University, Dnepropetrovsk 049050, Ukraine}} 
\date{}
\maketitle

\begin{abstract}
A new variational basis with well-behaved local approximation properties
and multiple output is
proposed for Coulomb systems.  
The trial function has proper behaviour at all Coulomb 
centres. Nonlinear asymptotic parameters are introduced softly: they do 
not destroy the self-optimized
local behaviour of the wave function at vanishing interparticle distances. 
The diagonalization of the Hamiltonian on a finite Hilbert subspace gives
a number of meaningful eigenvalues. Thus 
together with the ground state some excited states are also reliably 
approximated. For three-body systems all matrix elements are
analytically obtainable up to rational functions of 
asymptotic parameters. The feasibility of the new basis usage has been
proved by a pilot  computer algebra calculation. The negative sign of
an electron pair local energy at their Coulomb centre has been revealed.

\end{abstract}

\section{Introduction}
  In variational methods the required energy eigenvalue
is obtained  by averaging the Hamiltonian, or, in other terms, by averaging
the local energy, defined by action of the Hamiltonian operator on a trial wave
function. The very variability of the local energy is a direct consequence of
unavoidable approximate character of any chosen variational wave function. As
a result, the global (averaged) quantities are
much better reproduced in  variational  calculations than the local ones.
Moreover, in practical  calculations, the local reliability may be sacrificed
in favour of faster  convergence to the global values sought for. The known
notion of the effective charge $Z^{*}=Z-5/16$ ~in a two-electron atom (ion) is
a good example of such sacrifice, since the local
behaviour of an exact wave function
in the Coulomb  centre is determined by $Z$ unscreened but not by $Z^*$.

  The good local approximation quality of a variational wave
function is a highly desirable goal since the problem onset  
({\it Criteria of goodness for approximate wave functions}~\cite{James})
up until the present ({\it Quality of variational trial states}~\cite{Lucha}). 
For any Coulomb system this  quality is extremely
vulnerable at the Coulomb singularity points. Without special care the
local energy infinities take place at the Coulomb singularities, destroying 
the desired picture of the uniform approximation~\cite{Bethe}. 
Luckily, the average Hamiltonian values are rather insensitive
 to these local infinities due to: (1) much higher degree of smallness of the 
neighbouring integrated volume ($\sim r^{3}$) in comparison with the
 degree of the Coulomb singularity  ($\sim r^{-1}$) and (2) neutralizing of 
contributions with different signs at different points, as in customary cases
of  an effective charge or scale factor introduction. 

In spite of great 
achievements of modern variational calculations in finding high precision
energy values~\cite{Drake,Goldman,Frolov}, the
 problem of the reliable local approximation of the wave functions persists.
From now onwards, for brevity, the attribute ``local" will be related only to 
the Coulomb centres vicinities. 
 The Kato cusp conditions~\cite{Kato} 
 can be imposed as a supplementary condition for mean energy minimum,
 thus introducing conditional extremum technique, usually more laborious.
    The method not affecting variational freedom and, nevertheless, avoiding
 local energy infinities was proposed  earlier in~\cite{Vanyashin}. 
 The development of this method, presented below, adds the possibility to
 reproduce both local and asymptotic properties of an 
 exact solution in the basis functions, thus striving for better uniformity
of the wave function  approximation.

\section{
Coulomb Variational Basis with Both\\ Local and Asymptotic Proper Behaviour}
The mentioned local behaviour of a many-body wave function is, in
essence,  that of some superposition of Coulomb solutions for the
corresponding pair of  particles. In the case of an isolated Coulomb pair, the
wave function is a  well-known product of a normalization factor, radial, and
angular functions: \begin{eqnarray}
\psi =N\;R(\rho )\;Y({\bf n}) ;\;
\rho =-\frac{Z_1\; Z_2\;e^2\;m_1\;m_2}{\hbar ^2(m_1+m_2)}|{\bf
r}_1-{\bf r}_2| , \;{\bf n}=\frac{{\bf r}_1-{\bf r}_2}{| {\bf
r}_1-{\bf r}_2|},\;\nonumber\\ 
R(\rho )=\exp (-\rho /n)\;\rho ^l\; \Phi (1+l-n,\,2+2l;\,2\rho /n). 
\end{eqnarray}
The standard angular functions $Y({\bf n})$ can be equivalently
represented
as symmetric irreducible tensors of the rank $l$, composed  from the
Cartesian projections of the unit vector ${\bf n}$ and
Kronecker{\large$^{\bf ,}$}s deltas: \begin{eqnarray}
  \,l=0,\;Y=1;\;\;l=1,\;Y_i=n_i;\;\;
  l=2,\;Y_{ij}=3\;n_i\;n_j-\delta _{ij}; \nonumber\\
  l=3,\;Y_{ijk}=5\;n_i\;n_j\;n_k\;-\;\delta _{ij}\;n_k\;
-\;\delta_ {ik}\;n_j\;-\;\delta _{jk}\;n_i;\;\ldots   
\end{eqnarray}

The written Coulomb solution will be used not only for attracting pairs,
when it leads to the discrete spectrum of bound states, but also for
repulsing pairs. In the latter case the sequence of integral principal 
quantum numbers $n$ gives the corresponding sequence of the
Hamiltonian discrete  "eigenvalues". They are not physically meaningful for
isolated pairs: the "eigenfunctions" grow exponentionally and are not
normalizable. For repulsive pairs, which are embedded in a bound system, the
negative Hamiltonian  "eigenvalues" aquire the meaning of their local energies
near the Coulomb centre, as the exponentional growth of the "eigenfunctions"
will be damped by the environment.

  All products of the Coulomb wave functions of all  pairs, attracting and
 repulsing as well, can constitute the variational basis~\cite{Vanyashin}. 
A necessary contraction on dumb indices and selection of admissible
asymptotics  are implied. The permutation symmetry of identical particles
should be imposed on the  final form of the basis.

  Such a basis, put in order by integer principal and orbital quantum numbers
of  different pairs, is full enough to approximate any analytical many-body 
wave function. The Kato cusp conditions are rigorously satisfied by the basis 
functions themselves. We stress the point that any approximate fulfillment of
 the Kato cusp conditions leaves the difficulty of local energy infinities 
unsettled, no matter what precision of averaged quantities has been achieved.

  For any pair of particles from a many-body system only a density matrix can 
 be attributed and not a wave function. In the density matrix, unlike in the
 wave function of an isolated pair, the local and asymptotic parameters cannot
remain identical.
In order to  reproduce in the basis functions this density matrix
 property we modify the Coulomb radial functions so as to allow independent
 adjustment of local and asymptotic variational parameters.
 The modified radial function $R(a,\,\rho )$ is defined as the
 product of the exponential factor $\exp (-a\,\rho )$  and a finite
 segment of the Maclaurin expansion of the ratio of the unmodified  function
 $R(\rho )$ to the same exponential factor.  The modified
 function has two adjustable parameters: $a$ for the
 asymptotic behaviour, and $n$, real or imaginary, for the
 local behaviour near the Coulomb centre. This local behaviour is not affected
 by the performed soft introduction of the independent asymptotic behaviour,
which is not connected with that of a confluent hypergeometric function.
 The Maclaurin series of $\exp (a\,\rho )\,R(\rho )$
 up to $\rho ^{1+l+2k}$ appears to be the
 Laurent series in inverse even powers of $n$ up to $n^{-2k}$. Just so the
modified function has also the rearranged form:
\begin{equation}
R(a,\,\rho )=\sum _{k=0}^{k_{max}} \frac{c_k(a,\rho )}{n^{2k}}.
\end{equation}

   The Laurent coefficients $c_k(a,\rho )$  are proposed as the new
two-body  constituents of the many-body variational basis. They are
 independent of $n$ polynomials in both $\rho $ and $a$ with the common 
 exponential factor $\exp (-a\,\rho )$.
Along with stretching the basis set, the usage of $c_k(a,\rho )$
instead of $R(a,\,\rho )$ will absorb  the nonlinear parameters $n$
in easily obtainable coefficients of a linear superposition. 
The proposed
basis has inseparable cluster structure and the variational wave function 
should terminate only at the end of a cluster. In case of  three-body Coulomb
systems both the Hamiltonian and unity matrix elements can be computed
analytically up to rational functions of all asymptotic parameters
$a$.   

The squares of effective principal quantum numbers of all attractive and
repulsive Coulomb pairs, being included in the superposition coefficients,
are tuned automatically with the latters.    
So the proposed basis produces the multiple output:
not only the lowest root of the secular equation has the physical meaning,
but some higher roots are also meaningful.
Still a majority of higher roots remains  a mathematical artefact,
hence,  from the physical point of view, the Hamiltonian diagonalizes only
partially. 

\section{Computer Algebra Feasibility of the New Method } 
 The pilot computation  confirms the possibility of cooperative
treatment of several lowest  states. 
With a relatively short wave function containing 54 terms 
(49 terms in the antisymmetric
case) the helium para-S and ortho-S  energy levels  have been calculated in
one run as: \\
para-S levels: \ -2.902900, -2.145871, -2.055637;\\
ortho-S levels: -2.175026, -2.068634, -2.036463,\\ 
while the results of  high precision calculations
given in~\cite{Drake} are:\\ 
para-S levels: \ -2.903724377033982, -2.145974046054634,
-2.06127198974091;\\   ortho-S levels: -2.175229378237014, -2.068689067472468,
-2.03651208309830.\\   
Though these pilot numerical results are far from the record accuracy, they
have definitely  established the negative sign of the electron pair local
energy in the Helium atom. This phenomenon can be tested experimentally in,
{\it e. g.}, the Helium double ionization. The universal {\it Mathematica}
program {\it vSlevels} is available upon request from the author. 

  In our approach interparticle (Hylleraas) variables
are used for the basis  formation, that is natural for elimination of the
local energy infinities, and perimetric (Heron) variables --- on the stage of
analytical evaluation of  integrals, that simplifies calculations. In
hyperspherical variables the local energy infinities{\large$^{\bf,}$} problem 
appears more intricate. It has been solved principally through
frame transformations in the recent work~\cite{Heim}. 
\section*{Acknowledgments}

The author would like to thank V.~B.~Belyaev, 
B.~V.~Geshkenbein,\\ L.~B.~Okun{\large$^{\bf ,}$}, V.~S.~Popov, and  
Yu.~A.~Simonov for constructive discussions and valuable  comments.


\end{document}